\documentclass[aps,pra,twocolumn,groupedaddress]{revtex4-1}
\usepackage{graphicx}
\usepackage{amssymb}
\usepackage{amsmath}
\usepackage{mathbbol}
\usepackage{color}
\usepackage{textcomp}
\usepackage{wasysym}

\usepackage[utf8]{inputenc}
\usepackage{graphicx}
\usepackage{amssymb,amsmath}
\usepackage{tikz}
\usepackage{physics}
\usepackage{dutchcal}

\newcommand{\mf}{\mathcal{}}

\newcommand{\rn}{\textcolor{blue}}

\usepackage[colorlinks]{hyperref}

\def\({\left(}
\def\){\right)}

\def\av#1{\left\langle #1\right\rangle}

\def\a2d{a^{\dagger 2}}
\def\b2d{b^{\dagger 2}}

\def\eq#1{Eq.~(\ref{eq:#1})}

\def\fig#1{Fig.\ref{fig:#1}}

\def\tab#1{Table~\ref{tab:#1}}

\begin{document}
\title{Heisenberg-limited quantum interferometry with multi-photon subtracted twin beams} 

\author{Miller Eaton$^{1}$}
\email[]{me3nq@virginia.edu}
\author{Rajveer Nehra$^{1\dag}$}
\author{Aye Win$^{1\ddag}$}
\author{Olivier Pfister$^{1}$}
\affiliation{ ${}^1$Department of Physics, University of Virginia, 382 McCormick Rd, Charlottesville, VA 22903}
\affiliation{${}^\dag$Current Address: Department of Electrical Engineering, California Institute of Technology, Pasadena, California 91125, USA}
\affiliation{${}^\ddag$Current Address: Department of Physics, Old Dominion University, 4600 Elkhorn Ave
Norfolk, VA 23529}
\date{\today}

\begin{abstract}
We propose a Heisenberg-limited quantum interferometer whose input is twin optical beams from which one or more photons have been indistinguishably subtracted. Such an interferometer can yield Heisenberg-limited performance while at the same time giving a direct fringe reading, unlike for the twin-beam input of the Holland-Burnett interferometer. We propose a feasible experimental realization using a photon-number correlated source, such as non-degenerate parametric down-conversion, and perform realistic analyses of performance in the presence of loss and detector inefficiency.
\end{abstract}

\maketitle

\section{Introduction}

\begin{figure}[h!]
\centerline{\includegraphics[width=0.8\columnwidth]{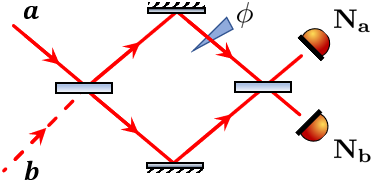}}
\caption{\em A Mach-Zehnder interferometer with phase difference $\phi$ between two optical paths. Both beamsplitters are balanced. Quantum splitting of input field $a$ implies interference with the vacuum field $b$. }
\label{fig:mz}
\end{figure}

A general interferometer, typified by the Mach-Zehnder interferometer (MZI) of \fig{mz}, measures the phase difference between two propagation paths by probing them with mutually coherent waves. From a purely undulatory standpoint, a sure way of ensuring such mutual coherence is to split an initial wave into  two  waves, for example by use of a beamsplitter. However, the unitarity of quantum evolution mandates that any two-wave-output unitary have a two-mode input as well --- rather than a classical, single-mode input.  Thus, the quantum description of a ``classical'' interferometer must feature an ``idle'' vacuum field in addition to the initial wave, and 
the quantum fundamental limit of interferometric measurements is then dictated by the corpuscular statistics of the interference between the two inputs of the beamsplitter (\fig{mz}).  In a classical interferometer, the vacuum fluctuations at the idle input port limit the phase difference sensitivity between the two interferometer arms to the quantum limit of classical interferometry~\cite{Caves1980},  the input beamsplitter's shot-noise limit (SNL)\footnote{The SNL is often called ``standard quantum limit.'' However, the latter was initially defined with a different meaning, in order to address the optimum error of quantum measurements in the presence of back action, such as radiation pressure on interferometer mirrors.~\protect\cite{Caves1980a,Braginsky}.}
\begin{equation}\label{eq:cl}
\Delta\phi_{SN} \sim \av N^{-\frac12}, 
\end{equation}
where $\phi$ is the phase difference to be measured and $N=N_{a}+N_{b}$ is the total photon number operator. This limit is that of phase noise {\em inside the interferometer} and has nothing to do with, say, the single-mode properties of a coherent state (e.g., laser) input $\ket\alpha$ of photon-number deviation $\Delta N = |\alpha|=\av N^{1/2}$ and phase deviation~\footnote{From the number-phase Heisenberg inequality $\Delta N\Delta\phi\geqslant1/2$, easily derived from the energy-time inequality~\cite{JMLL}.} $\Delta\theta \sim \av N^{-1/2}$ {\em before the interferometer}. In fact, Caves showed that a Fock-state input $\ket n$, for which $\Delta N=0$ and hence $\Delta\theta\to\infty$, still yields the SNL of \eq{cl}~\cite{Caves1980}.  

When both input modes of the interferometer are properly ``quantum engineered,'' one can, in principle, reach the ultimate limit, called the Heisenberg limit (HL), 
\begin{equation}\label{eq:hl}
\Delta\phi_{H} \sim \av N^{-1},
\end{equation}
which can clearly be many orders of magnitude lower than the SNL when $\av N\gg1$. A recent comprehensive review of quantum interferometry can be found in Ref.~\citenum{Demkowicz2015}. The first quantum engineering proposal to break through the SNL  was Caves'  idea to replace the vacuum state input with a squeezed vacuum~\cite{Caves1981}, which has since been shown to optimize the quantum Cram\'er-Rao bound when the input field is a coherent state~\cite{Lang2013}. This was demonstrated experimentally~\cite{Grangier1987,Xiao1987} and is now the approach adopted for high-frequency signals (above the standard quantum limit) in gravitational-wave detectors~\cite{LIGO2011,Aasi2013}. Many other approaches have been investigated~\cite{Yurke1986a,Luis2000}, such as twin beams~\cite{Holland1993,Bouyer1997,Kim1998,Dowling1998,Feng2004,Feng2004a}, N00N states~\cite{Bollinger1996,Boto2000,Mitchell2004,Walther2004,Afek2010}, or two-mode squeezed states. These different schemes were recently compared in terms of the quantum Cram\'er-Rao bound on their phase sensitivity~\cite{Lang2014}.  

It is important to recall here the essential result of Escher, de Matos Filho, and Davidovich: operating a realistic, i.e., lossy, interferometer at the HL requires losses to be no greater than $\av N^{-1}$~\cite{Escher2011}, i.e., the grand total of the loss can never exceed one photon, on average. This result had been obtained earlier by Pooser and Pfister in the particular case of Holland-Burnett interferometry~\cite{Pooser2004}: using Monte Carlo simulations for up to $n=10\,000$ photons, it was shown that the phase error of a non-ideal Holland-Burnett interferometer scales with the HL if the losses are of the order of $n^{-1}$, and that larger losses degrade the scaling to a limit proportional to the SNL of $N^{-1/2}$,  staying sub-SNL as long as photon correlations are present in the twin Fock input.  This is consistent with the general result of Escher, de Matos Filho, and Davidovich for phase estimation~\cite{Escher2011}. Additionally, in the case of loss, optimal input states have been numerically calculated whose form generally depends on the interferometric efficiency and average photon-number input ~\cite{Dorner2009optimal}.

A direct consequence is that, if the total photon number is too large, ultimate-sensitivity interferometry cannot be Heisenberg-limited in the current state of technology: the most sensitive classical interferometer to date, the Laser Interferometer Gravitational-wave Observatory (LIGO) before the introduction of squeezed light boasted $\Delta\phi_{SN} \sim 10^{-11}$ rad and is shot-noise-limited in some spectral regions, therefore featuring $\av N\sim 10^{22}$ photons. While a Heisenberg-limited version of  LIGO would only require $\av N\sim 10^{11}$ photons to reach the same sensitivity, it would also require an unrealistic loss level of $10^{-11}$, the optical coatings on LIGO's mirrors ``only'' reaching already remarkable sub-ppm loss levels~\cite{aLIGO2014}. 

However, the maximally efficient use of photons by Heisenberg-limited interferometry can still be interesting provided we take into account this constraint of an ultimate loss level of $10^{-6}$. At this level, a 1064 nm interferometer with (arbitrarily chosen) 10 ms measurements would be allowed to reach the $10^{6}$-photon HL of 1 $\mu$rad with only 200 pW, whereas a classical interferometer would need $10^{12}$ photons, i.e., 200 $\mu$W, to have its SNL at 1 $\mu$rad. This can be of interest in situations where low light levels are beneficial, such as phase imaging of living biological tissue.

In order to motivate the approach of this paper, we review and compare and contrast some different quantum-enhanced sensing proposals in \tab p. The key points we examine are: \\
{\em(i)}, whether the input state enables performance at the HL;\\ 
{\em(ii)}, whether a direct interference fringe is observable;\\ 
{\em(iii)}, whether the  $\av N\gg1$ regime is experimentally accessible. As will be shown, the new input state we propose in this paper is the only one that fulfills all three criteria. 
\begin{widetext}

\begin{table*}[h!]
\caption{Characteristics and performance of different input states --- except the N00N state$^{*}$ which is a state specified {\em inside} the interferometer. Also the fringe signal for the N00N state requires $n$-photon detection$^{(\ddag)}$. The phase error $\Delta\phi$ is the quantum Cram\'er-Rao bound~\cite{Lang2014}. The state whose use we propose in this paper is  the last one.}
\begin{center}
\begin{tabular}{|c|@{\quad }c@{\quad }|c|cc|c|c|}
\multicolumn{2}{c|}{Input$^{*}$ state} &Ref.& {\em (i)} $\Delta\phi$&  & {\em (ii)}fringe $\av{N_{a}-N_{b}}$ & {\em (iii)} $\av N\gg1$? \\
\hline
&&&&&&\\
1. &$\ket n_{a}\ket0_{b}$&\cite{Caves1980} & $\displaystyle\frac1{\sqrt n}$ & SNL& $n\cos\phi$ & yes \\
&&&&&&\\
2. &$\ket\alpha_{a}\ket0_{b}$&\cite{Caves1980} & $\displaystyle\frac1{\sqrt{\av{N}}}=\frac1{|\alpha|}$ &SNL& $|\alpha|^{2}\cos\phi$ & yes \\
&&&&&&\\
3. &$\ket\alpha_{a}\ket{0,r}_{b}$&\cite{Caves1981} & $\displaystyle\simeq\frac{e^{-r}}{|\alpha|}$ &sub-SNL& $|\alpha|^{2}\cos\phi$ & yes \\
&&&&&&\\
4. &$\ket n_{a}\ket n_{b}$
&\cite{Holland1993} & $\displaystyle\frac1{\sqrt{2n(n+1)}}$ 
&HL& 0 & \qquad yes~\cite{Feng2004} \\
&&&&&&\\
5. &$\ket n_{a}\ket{n-1}_{b}$
&\cite{Lang2014} & $\displaystyle\frac1{\sqrt{2n^2-1}}$ &HL& $\displaystyle\frac12\cos\phi$  & possible \\
&&&&&&\\
$6.^{*}$ &$\displaystyle\frac1{\sqrt2}(\ket n_{a}\ket0_{b}+\ket0_{a}\ket n_{b})$
&\cite{Bollinger1996,Boto2000} & $\displaystyle\frac1{n}$&HL & $\sim\cos(n\phi)^{(\ddag)}$ & unknown \\
&&&&&&\\

7. &$\displaystyle\frac1{\sqrt2}(\ket n_{a}\ket n_{b}+\ket{n+1}_{a}\ket{n-1}_{b})$&\cite{Yurke1986a} & $\displaystyle\frac{1}{\sqrt{2n(n+1)-1}}$& HL & $\frac{\cos\phi}2-\frac{\sin\phi}4\sqrt{n(n+2)}$ & unknown \\

&&&&&&\\

8. &$\displaystyle \sum_{k=0}^Z c_k \ket{n-Z+k}_a\ket{n-k}_b$ && $\displaystyle \simeq \frac{\sqrt{Z}}{n}$ &HL& $\displaystyle \sim\frac n2\sin\phi$ &  possible \\
&&&&&&\\
\hline
\end{tabular}
\end{center}
\label{tab:p}
\end{table*}
\end{widetext}

The first two cases are interferometry with the vacuum field in one beamsplitter input port, leading to no quantum enhancement.

The third case makes use of Caves' squeezed input~\cite{Caves1981} into the previously unused port of the beamsplitter. This benefits from mature, high-level laser and quantum optics technology, with large average photon numbers from well stabilized lasers~\cite{Drever1983}. Case 3 benefits from the recent 15 dB squeezing record~\cite{Vahlbruch2016}, but it does require that  the  phase difference between the squeezed state and the coherent state be controlled~\cite{Vahlbruch2006}. The gravitational-wave observatories of Advanced LIGO, Advanced VIRGO, and GEO600 all currently utilize squeezed light to improve sensitivity~\cite{tse2019quantum, acernese2019increasing, grote2013first}, and Advanced LIGO will soon implement frequency-dependent squeezing to improve sensitivity over a larger bandwidth~\cite{mcculler2020frequency}.

Case 4 in \tab p is the twin Fock state input first proposed by Holland and Burnett~\cite{Holland1993}, and which is implementable, to a good approximation, with large photon numbers by using an optical parametric oscillator above threshold~\cite{Heidmann1987,Mertz1991,Laurat2005,Feng2004,Feng2004a}. The input density operator is then of the form, in the absence of losses,
\begin{equation}
\rho = \sum_{n,n'} \rho_{n,n'}\ket{n n}\bra{n' n'},
\end{equation}
which can be a pure state ($ \rho_{n,n'}\mapsto\rho_{n}\rho_{n'}^{*}$), e.g.\ the two-mode squeezed state emitted by a lossless optical parametric oscillator (OPO) below threshold, or can be a general statistical mixture as emitted by a lossless OPO above threshold~\footnote{This is the most general mixture for which $\protect\av{N_a-N_b}=0$ and $\protect\Delta({N_a-N_b})=0$.}. It thus also benefits from the same mature OPO-based squeezing technology, with a record 9.7 dB reduction on the intensity-difference noise~\cite{Laurat2005}. Moreover, the phase difference between the twin beams is irrelevant (being actually very noisy from being anti-squeezed) and thus need not be controlled before the interferometer. The generalized~\cite{Campos1989} Hong-Ou-Mandel~\cite{Hong1987} quantum interference responsible for twin beams breaking the SNL was demonstrated experimentally in an ultrastable phase-difference-locked OPO above threshold~\cite{Feng2004,Jing2006,Pysher2010a}, with several mW of CW power.

An inconvenient feature of the Holland-Burnett scheme, however, is that the direct interference fringe disappears ($\av{N_{a}-N_{b}}=0$ in \tab p, a property also shared by the classical input $\ket\alpha_{a}\ket\alpha_{b}$) in contrast to all previous cases for which the fringe signal is proportional to the total photon number. This inconvenience can be circumvented by the use of Bayesian reconstruction of the probability distribution~\cite{Holland1993,Kim1998,Pooser2004}. However, this requires photon-number-resolved detection at large photon numbers, which isn't  accessible experimentally yet. Another workaround is to use the variance of the photon-number difference, which is sensitive to $\phi$~\cite{Bouyer1997} but whose  signal-to-noise ratio is bounded by $\sqrt2$~\cite{Kim1998}. Another idea is to use a heterodyne signal, which presents high visibility but is restricted to phase shifts ever closer to zero as the squeezing increases~\cite{Snyder1990}. This was demonstrated experimentally as heterodyne polarimetry 4.8 dB below the SNL~\cite{Feng2004a}. 

Case 5 in \tab p is a variant of the twin Fock state, the ``fraternal'' twin Fock state~\cite{Lang2014}, which does provide a direct fringe signal which being Heisenberg-limited, but the fringe signal is still extremely small as it does not scale with the input photon number. Recently, other variants of this state in the family $|n\rangle_a|m\rangle_b$ have been shown better performance than Holland-Burnett and N00N states in the presence of loss~\cite{thekkadath2020quantum}.

Case 6 stands out for several reasons. The N00N state refers not to an interferometer input state but to a state inside the Mach-Zehnder interferometer~\cite{Bollinger1996,Boto2000}. While it yields performance at the HL, its experimental generation is experimentally inaccessible for $n\gg1$; previous experimental realizations have used post-selected outcomes for $n=3$~\cite{Mitchell2004} and 4~\cite{Walther2004}, a method which fails to scale up to large photon numbers, though a more scalable method using coherent state displacement was also demonstrated~\cite{Afek2010}. Furthermore, N00N states are extremely vulnerable to loss~\cite{kacprowicz2010experimental}, although experimental efforts have attempted to side-step the issue by using more loss-tolerant techniques~\cite{Ulanov2016} and ultra-high efficiency sources and detectors~\cite{slussarenko2017unconditional}. Last but not least, the use of a N00N state with $n$ photons requires $n$-photon detection, which is currently inaccessible in optics experiments with $n\gg1$ (but may be easier to reach in atomic spectroscopy~\cite{Bollinger1996}).

Case 7 is the theoretical proposal of Yurke, McCall, and Klauder (YMCK)~\cite{Yurke1986a}. It features both performance at the HL and a strong fringe signal, but an experimental realization has yet to be determined.

Case 8 features the input proposed in this paper; it is the only one of the table that features HL performance, a clear interference fringe signal, and is experimentally feasible with demonstrated technology for large photon numbers. The state can be generated by using bright twin beams from which one or multiple photons have been coherently subtracted.

In addition to the above cases, photon subtraction has been suggested to be used in other interferometry schemes, such as interfering a coherent state and photon-subtracted squeezed vacuum~\cite{birrittella2014quantum} or subtracting a photon from each mode of a two-mode squeezed state~\cite{Carranza2012} ({\em distinguishable} subtraction, unlike what we propose here), but these schemes would require either a parity measurement or large photon-number resolving (PNR) measurements for end-detection, both of which are currently unfeasible for large photon numbers. As will be demonstrated, our method can make use of bright twin beams, thus scaling to large numbers of photons, and provides a directly measurable fringe with conventional photo-diodes.

This paper is organized as follows.  In section~\ref{sec:fisher}, we introduce the Schwinger-spin representation and calculate the Cram\'er-Rao bound by way of the quantum Fisher information for our proposed state. Section~\ref{sec:MZI} demonstrates that the $Z$-photon coherently-subtracted twin-beam state has phase-sensitivity scaling with the HL when implemented with a traditional MZI. We then propose an experimental scheme to generate the desired state in section~\ref{sec:experiment} and derive the result from a twin-beam input. Section~\ref{sec:sqz_input} shows the results of numerical calculations when all approximations are disregarded, and section~\ref{sec:practical} discusses practical considerations of loss, detection imperfections, and the use of click detectors in place of PNR detectors during state generation. 

\section{Quantum Fisher Information} \label{sec:fisher}
The Fisher information is a well-known parameter that provides a means to quantify the amount of information contained by a parameterized random variable, and the Cram\'er-Rao inequality formulates an upper bound on the precision of an estimator in terms of the Fisher information~\cite{Cramer1999mathematical,Rao1992information}. This inequality has been extended to the quantum case~\cite{helstrom1969quantum} which is very useful in determining the bounds on the sensitivity for quantum interferometry given a specified input state~\cite{Lang2013}, and is independent of the nature of the estimator. For a standard MZI, the estimator of interest is the phase difference between the two arms of the interferometer, for which the quantum Fisher information of a general pure input state is given by~\cite{jarzyna2012quantum,Lang2013}
\begin{equation}
    \mathcal{F}=\bra{\psi_{in}}U_{BS}^\dag N_d^2 U_{BS}\ket{\psi_{in}} - \bra{\psi_{in}}U_{BS}^\dag N_d U_{BS}\ket{\psi_{in}}^2,
\end{equation}
where $U_{BS}=exp(\frac{i\pi}{4}(a^\dag b + ab^\dag))$ is a balanced beamsplitter and $N_d=a^\dag a-b^\dag b$ is the photon number difference operator. For the sake of convenience, we adopt for our calculation the Schwinger-spin SU(2) representation~\cite{Schwinger1965} initially proposed by Yurke et al.\ for quantum interferometers~\cite{Yurke1986a}. A fictitious spin $\vec J$ is defined from a pair of bosonic modes $(a,b)$ as
\begin{align}
    J_z&=\frac12(a^\dag a - b^\dag b)\\
    J_x&=\frac12(a^\dag b + a b^\dag)\\
    J_y&=-\frac{i}{2}(a^\dag b - a b^\dag),
\end{align}
where ${a}$ and ${b}$ are the annihilation operators of each mode. These operators satisfy the canonical angular momentum commutation relations of the SU(2) algebra
\begin{equation}
[{J_i}, {J_j}] = i \epsilon_{ijk}{J_k}.
\end{equation}
The operator ${J_z}$ represents the photon number difference operator between the two modes whereas ${J_{x,y}}$ are interference terms. The ease of working with Schwinger operators is further simplified by noting that the common eigenstates of $J_z$ and $J^2$, $\ket{j,m}$, take the form of two-mode Fock states:
\begin{equation}
    \label{eq:jm}
    \ket{j\rn{,}m}=\ket{n_a}_a\ket{n_b}_b,
\end{equation}
where we have
\begin{align}
    j&=\frac12 (n_a+n_b)\text{; }  \\
    m&=\frac12 (n_a-n_b).
    \label{eq:m}
\end{align}
We can then see that the Fisher information can be expressed as
\begin{align}
    \mathcal{F}=\,&4\bra{\psi_{in}}e^{(-i\frac{\pi}{2}J_x)}J_z^2 e^{(i\frac{\pi}{2}J_x)}\ket{\psi_{in}} \nonumber\\
    &- 4\bra{\psi_{in}}e^{(-i\frac{\pi}{2}J_x)} J_z e^{(i\frac{\pi}{2}J_x)}\ket{\psi_{in}}^2\\
    =\,&4\bra{\psi_{in}}J_y^2\ket{\psi_{in}} - 4\bra{\psi_{in}}J_y\ket{\psi_{in}}^2\\
    =\,&-\langle(a^\dag b -a b^\dag)^2\rangle + \langle a^\dag b - a b^\dag\rangle^2.
    \label{eq:fisher}
\end{align}
This quantity is related to the phase estimation bound by
\begin{equation}
 (\Delta \phi_d)^2\geq \frac{1}{\mathcal{F}}.
\end{equation}
It is important to note that $\Delta \phi_d$ is the general phase difference measurement whereas the quantum Fisher-limited phase error, i.e., when the inequality achieves equality, can be achieved for the correct estimator when $\phi$ deviates from an initially specified optimal value.

For an interferometry scheme using solely a coherent state input, the Fisher information can be easily calculated to yield exactly the expectation value of the input photon number, $\mathcal{F}=\langle N\rangle$. This leads to a maximum phase sensitivity of $\Delta \phi_d=\langle N\rangle^{-1/2}$, which is the well-known limit to interferometric measurement due to quantum noise with a classical input shown in Table~\ref{tab:p}, Case 2. In order to beat this limit, it becomes necessary to include something other than coherent states and vacuum as input, such as squeezed light addressed by Case 3 of Table~\ref{tab:p}. Although this will beat the $\langle N\rangle^{-1/2}$ scaling, it still cannot achieve the HL and has a maximum scaling of $\sim \langle N\rangle^{-2/3}$, which is only achievable with large squeezing ($\sim |\alpha|^{2/3}$ photons in the squeezed field for $|\alpha|^2$ photons in the coherent state)~\cite{Caves1980}. Because of this, reaching the HL requires the use of other types of quantum states.
As a start, consider the simplest case of the final state in \tab p, (Z=1 case), the state given by
\begin{equation}
    \ket{\psi^{(i)}}=A\ket{n-1}_a\ket{n}_b+B\ket{n}_a\ket{n-1}_b,
    \label{eq:PSTB}
\end{equation}
where $|A|^2+|B|^2=1$. Calculating the quantum Fisher information for $\psi^{(i)}$ in Eq.~\ref{eq:PSTB}, gives 
\begin{equation}
    \mathcal{F}^{(i)}=2n^2-4n^2\text{Im}[A^*B]-1,
\end{equation}
which leads to the inequality
\begin{equation}
    (\Delta \phi^{(i)}_d)^2\geq\frac{1}{2n^2-1}
    \label{eq:Dphi_i}
\end{equation}
when $A=B=\frac{1}{\sqrt{2}}$. In general, it is possible to achieve the Cram\'er-Rao bound by judiciously choosing the ideal measurement scheme~\cite{braunstein1994statistical}. If the ideal measurement is physically realizable, such as photon-number difference or single-mode parity, Eq.~\ref{eq:Dphi_i} demonstrates that the maximum sensitivity of $\Delta \phi_d$ can achieve a Heisenberg-limited scaling of $\langle N \rangle ^{-1}$.

We now consider the most general state with arbitrary $Z$:
\begin{align}
    \ket{\psi}=&c_0\ket{n-Z}_a\ket{n}_b+c_1\ket{n-Z+1}_a\ket{n-1}_b \nonumber\\
    &+\hdots+c_Z\ket{n}_a\ket{n-Z}_b \nonumber \\
    =&\sum_{k=0}^Z c_k \ket{n-Z+k}_a\ket{n-k}_b,
    \label{eq:gen_coh_subtraction}
\end{align}
where some initial $\ket{n}_a\ket{n}_b$ has been subjected to a coherent $Z$ photon subtraction, where the $Z$ photons could have been removed from either mode in any given combination. Derived in Appx.~\ref{appx:gen_fisher}, the Fisher information when all $c_k$ are the same and the mean photon number of the state is much larger than the number of photons subtracted is determined to be 
\begin{equation}
    \mathcal{F}\stackrel{(i)}{=}\frac{N^2}{Z},
\end{equation}
where $(i)$ represents the use of the condition that $N=2n-Z \gg Z$. This leads to a bound on the phase sensitivity to be
\begin{equation}
    (\Delta \phi_d)^2\geq\frac{Z}{N^2},
\end{equation}
which shows the potential to achieve scaling that is proportional to HL. Although the bound given by the Fisher information is general, true infererometric performance is dependent upon the measurement scheme. We will now demonstrate that the class of states given by Eq.~\ref{eq:gen_coh_subtraction} follows the $N^{-1}$ scaling in a realistic MZI implementation when subtracting the resultant photo-detection currents, and will also yield a measurable interference fringe scaling with input photon number. From there, we will provide a feasible experimental scheme based on correlated twin-beams and a single multi-photon subtraction event.

\section{MZI implementation sensitivity}\label{sec:MZI}
The above discussion demonstrated that the generalized photon-subtracted states are potentially viable for Heisenberg-limited interferometry as verified by the quantum Fisher information, but now it is important to ensure that the same $\langle N \rangle^{-1}$ scaling can be reached with a realistic implementation, such as subtracting photocurrents, and that there is a measurable fringe \textit{also} scaling with $\langle N \rangle$ to ensure a sufficient signal above any electronics noise floor. The standard MZI is shown for reference in Fig.~\ref{fig:mz}, where $\Delta \phi$ is the phase difference between the arms of the interferometer and the end detectors measured the respective mean photocurrents, $N_a$ and $N_b$.
The input state, $\ket{\psi}_{ab}$, is transformed using the Schr\"odinger Picture by the interferometer to 
\begin{equation}
    \ket{\psi'}_{ab}=U^\dag_{BS}P_{\Delta \phi}U_{BS}\ket{\psi}_{ab},
\end{equation}
where $P_{\Delta \phi}$ applies the relative phase shift $\Delta \phi$. This is equivalent, in the Schwinger representation, to applying the rotations
\begin{equation}
    \ket{\psi'}_{ab}=e^{-i\frac{\pi}{2}J_x}e^{i\phi J_z}e^{i\frac{\pi}{2}J_x}\ket{\psi}_{ab}.
\end{equation}
However, in order to determine if there is a measurable fringe, it is necessary to find $\langle N_a-N_b\rangle$, which is much easier to find when working in the Heisenberg Picture, as it simply corresponds to the transformed rotation operator, $\langle J_z' \rangle$, where any operator, $\mf{O}$, is transformed by the MZI to become
\begin{equation}
    \mf{O'}=e^{-i\frac{\pi}{2}J_x}e^{-i\phi J_z}e^{i\frac{\pi}{2}J_x}\mf{O}e^{-i\frac{\pi}{2}J_x}e^{i\phi J_z}e^{i\frac{\pi}{2}J_x}.
\end{equation}
Using the transformation, we find
\begin{equation}
    J_z'=\cos{\phi}J_z-\sin{\phi}J_x.
    \label{eq:jz}
\end{equation}
Making use of Eq.~\ref{eq:jm}-\ref{eq:m}, we can rewrite the state of interest from Eq.~\ref{eq:gen_coh_subtraction} in the Schwinger representation to be
\begin{equation}
    \ket{\psi}=\sum_{m=-s}^{s}c_m\ket{jm},
    \label{eq:gen_coh_schw}
\end{equation}
where $s=\tfrac12 Z$, $j=n-s$, and from this point forward we consider all coefficients real and symmetric such that
\begin{equation}
    \label{eq:coef_condition}
    c_m^*=c_m=c_{-m},
\end{equation} 
which is the case for the physical state as derived in Appx.~\ref{appx:exp_deriv}. Supposing condition $(i)$ that the number of photons removed from the state by the subtraction process is considerably smaller than the number of photons remaining, i.e. $j\gg s$, the measurable fringe for the state in Eq.~\ref{eq:gen_coh_schw} is given by
\begin{align}
    \frac{j}{s}|\sin{\phi}| &\stackrel{(i,ii)}{\leq} 2|\langle J_z'\rangle| \stackrel{(i)}{\leq} 2j|\sin{\phi}|,
    \label{eq:fringe}
\end{align}
with $(ii)$ denoting the use of results from Appx.~\ref{appx:coef} where we bound the closeness of neighboring coefficients $c_m, c_{m+1}$ obtained from the experimental design. Eq.~\ref{eq:fringe} shows that there is a direct measurable fringe scaling with $j$. Now, in order to determine the phase sensitivity when using $J_z'$ as a phase estimator, it is necessary to calculate the quantity
\begin{equation}
(\Delta \phi)^2=\left[\Delta J_z' \left(\frac{d}{d\phi}\langle J_z' \rangle\right)^{-1}\right]^2.
\label{eq:dphi_MZI}
\end{equation}

\begin{figure*}[ht!]
    \centering
\includegraphics[width = 0.9\textwidth]{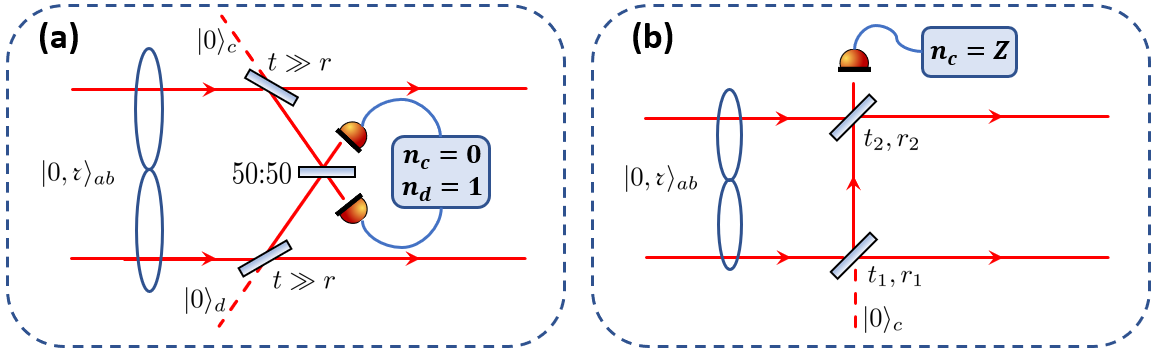}
\caption{(a) Scheme using two click detectors to indistinguishably subtract a single photon from two squeezed modes. (b) Simplified version using a single detector that includes the case of coherent single-photon subtraction, but also extends the method to include coherent $Z$-photon subtraction when a PNR detector is used.} 
    \label{fig:diagram}
\end{figure*}
Derived in Appx.~\ref{appx:MZI_calcs}, $\Delta\phi$ takes on a minimum value about the angle $\phi=0$, yielding an upper bound on the phase-sensitivity of
\begin{equation}
    \frac{\sqrt{s}}{j} \stackrel{(i)}{\leq} \Delta \phi_{min} \stackrel{(i,ii)}{\leq} \frac{2s^2}{j}.
    \label{eq:phi_min}
\end{equation}
We note again that $s$ is a small number based on the number of photons subtracted, so is simply a constant factor that does not affect scaling. Together, Eqs.~\ref{eq:fringe} and~\ref{eq:phi_min} show that the general $Z$-photon subtracted state has sensitivity proportional to the HL and has a measurable fringe that scales with photon number, meaning that this class of states is potentially useful for quantum-enhanced interferometry, provided a physical realization can be found. Such a realization is provided in the next section, along with a detailed analysis of the possible performance.

\section{State generation}\label{sec:experiment}

The specific case of $Z=1$ has been previously examined in Ref.~\cite{hofmann2006generation}, where the proposed experimental design is shown here in Fig.~\ref{fig:diagram}(a). By sending each mode of a two mode squeezed (TMS) state to a highly unbalanced beamsplitter, a single photon is subtracted from one of one of the modes.  However, before the detection occurs, a third balanced beamsplitter is cleverly placed to erase the identifying path information about from which mode the photon came.  By detecting exactly one photon on the combined two detection modes, the scheme implements a superposition of performing the subtraction on each mode to create a superposition of the type of states given in Eq.~\ref{eq:PSTB}, which has the form of
\begin{equation}
    \sum_n c_n (\ket{n}_a\ket{n-1}_b +\ket{n-1}_a\ket{n}_b),
\end{equation}
where the coefficients $c_n$ depend on the strength of the initial TMS state.

Here, we demonstrate that this state, along with all classes of state given by Eq.~\ref{eq:gen_coh_subtraction} can be generated using only a single detector as per the scheme shown in Fig.~\ref{fig:diagram}(b). In this implementation, instead of mixing both modes with vacuum to perform the subtraction, an asymmetric scheme is used where only one mode mixes with vacuum while the other is mixed with the siphoned-off portion of the first beam. By tuning the beamsplitter coefficients, it is possible to make a variety of interesting entangle states beyond the scope of Eq.~\ref{eq:gen_coh_subtraction}, but we will show that utilizing beamsplitters with high transmissivity will preserve higher mean photon numbers of the outputs for any given detected number of photons, $Z$.

The first step of Fig.~\ref{fig:diagram}(b) starts with a two-mode squeezed input state given by
\begin{equation}
    \ket{\phi_{in}}=\ket{0,\mathcal{r}}_{ab}\ket{0}_c=\frac{1}{\cosh{\mathcal{r}}}\sum_n^\infty (\tanh{\mathcal{r}})^n \ket{n}_a\ket{n}_b\ket{0}_c,
\end{equation}
where $\mathcal{r}$ the squeezing parameter. The state is sent to the beamsplitters each with real reflectivity and transmissivity coefficients such that $t_i^2+r_i^2=1$. Next,the mode $c$ is sent to a PNR detector to measure $Z$ photons and project the remaining modes into the desired state. The output state for the general case is derived in Appx.~\ref{appx:exp_deriv} and may be useful for engineering interesting two-mode quantum states, but the most desirable case for use in sensitive interferometric measurements occurs when both beamsplitters are highly transmissive such that $t\gg r$. 
This leads to the approximate output state, conditioned on a detection of $Z$ photons, to be
 \begin{align}
     \ket{\phi_{out}}\propto&\sum_{n=0}^\infty (t^2 \tanh{\mathcal{r}})^n  \sum_{k_{min}}^Z\sqrt{\binom{n}{k}\binom{n}{Z-k}\binom{Z}{k}}\nonumber \\
     &\times e^{ik\varphi}\ket{n+k-Z}_a\ket{n-k}_b,
     \label{eq:approx_state}
\end{align}
where $\varphi$ is the phase difference between the two input modes to the second beamsplitter{~($t_2$, $r_2)$} and $k_{min}=\text{Max}(0,Z-n)$. This state is simply a superposition of the general class of states given in Eq.~\ref{eq:gen_coh_subtraction}, where for each $\ket{nn}$ term, the $Z$ photons have been coherently subtracted from both modes in every possible configuration. When $\varphi$ is set to zero, the coefficients exactly follow the properties specified above in Eq.~\ref{eq:coef_condition}. An important point to note is that the value of $n$ in each input $\ket{nn}$ term is determine by the initial squeezing, but modified by the beamsplitter transmissivity to yield a new decreased effective squeezing, $\tanh{\mathcal{r}}\rightarrow t^2\tanh{\mathcal{r}}$.  Regardless of the initial squeezing value, this reduction in squeezing means that only values of $t\approx1$ will allow for useful interferometric measurements as large $n$ is desirable. Cases with $t$ deviating from unity may still work for interferometry applications if a large value of $Z$ is measured; however, this is prohibited by challenges in performing such a large PNR measurement.\\
\section{Squeezed Input}\label{sec:sqz_input}
Numerical simulations were performed using the Python package QuTip~\cite{Johansson2012} to determine the minimum resolvable fringe from an MZI according to Eq.~\ref{eq:dphi_MZI}, where the input state was a two-mode squeezed state that had undergone photon-subtraction as per Fig.~\ref{fig:diagram}b. 
\begin{figure}[ht]
\centering
\includegraphics[width = 0.99\columnwidth]{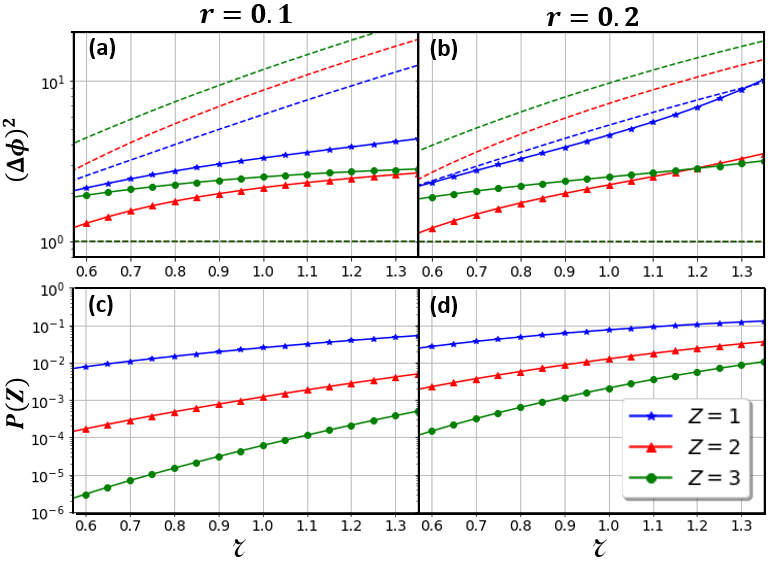}
\caption{Minimum phase sensitivity for a photon-subtracted state where a PNR detector has detected $Z$ photons during state generation where beamsplitter reflectivities are $r_1=r_2=0.1$ (a) and $r_1=r_2=0.2$ (b). States are normalized to their respective HL (lower dotted line) with upper dotted lines showing the respective shot-noise levels. Probability to successfully detect $Z$ photons vs input state squeezing for the beamsplitter coefficients $r_1=r_2=0.1$ (c) and $r_1=r_2=0.2$ (d) are also shown.}
    \label{fig:sig_phi}
\end{figure}

The results are shown in the top panel of Fig.~\ref{fig:sig_phi}, where $Z$ varies from one to three photons as the initial squeezing is increased, and both beamsplitters are taken to be equivalent with reflectivity $r_1=r_2\equiv r$.  All cases beat the shot-noise limit, but it is clear that decreasing the reflectivity leads to a larger quantum-enhancement, and larger values of $Z$ increase the overall phase-sensitivity. This behavior may seem odd, as looking at Eq.~\ref{eq:phi_min} shows that the overall scaling of the sensitivity is slightly worse as $Z$ increases; however, detecting a larger number of photons acts to increases the mean photon-number of the resultant state before the interferometer, and higher-order error terms from the use of beamsplitters with non-vanishing reflectivity drop off more quickly with increasing $Z$. For input states with lower mean-photon number, post-selecting on a larger $Z$ can actually beat the HL set by the states heralded at smaller $Z$. However, this comes at the cost of diminishing success rates. 

By comparing Fig.~\ref{fig:sig_phi}a and b, it is clear that for the relatively large value of $r=0.2$ the weak beamsplitter approximation employed earlier is no long valid for smaller $Z$, as the state with $Z=1$ is only marginally better than the classical case.  However, decreasing the reflectivity to $r=0.1$ suffices to give a quantum advantage that scales when increasing the input energy for all detected values of $Z$. This highlights the importance of choosing an optimal beamsplitter coefficient that pairs with the desired post-selection value, $Z$, to yield increased sensitivity while balancing success rates for a given input state.

\section{Practical considerations}\label{sec:practical}
\subsection{PNR Detector Loss}
In a realistic implementation, one might think that the state generation is highly sensitive to PNR detector efficiency.  However, provided any imperfections in the detector are not too drastic, the present scheme is tolerant to this inefficiency. Instead of projecting mode $c$ of the intermediate state following the beamsplitters onto an ideal measurement of $Z$ photons, consider a detector positive-operator valued measure (POVM) given by 
\begin{equation}
    D(Z)=\sum_{l=Z}^\infty \binom{l}{Z}\eta^Z (1-\eta)^{l-Z}\ket{l}\bra{l},
    \label{eq:povm}
\end{equation}
where the detector has efficiency $\eta$ and the device registers a detection of $Z$ photons. The output density matrix will be given by
\begin{align}
\rho&\propto\text{Tr}_c\left[D_c(Z)\left(\ket{\psi}\bra{\psi}\right)_{abc}\right],
\end{align}
where $\ket{\psi}$ is the state following both beamsplitters given by Eq.~\ref{eq:appx_state_pre_pnr}. This leads to 
\begin{equation}
    \rho\propto\sum_{l=Z}^\infty \binom{l}{Z}\eta^Z (1-\eta)^{l-Z}\rho_l,
\end{equation}
where $\rho_l$ is the density matrix of the pure state for an $l$-photon detection given in Eq.~\ref{eq:appx_genresult}. However, taking $r_1=r_2\equiv r$ and $r \ll 1$ yields
\begin{align}
    \rho&\approx\sum_{l=Z}^\infty \binom{l}{Z}\eta^Z (1-\eta)^{l-Z}\left(\frac{r}{t}\right)^{2l}\rho'_l\\
    &\propto\sum_{l=Z}^\infty \binom{l}{Z}\left(\frac{r^2(1-\eta)}{t^2}\right)^{l}\rho'_l,
    \label{eq:approx_povm_dm}
\end{align}
where $\rho'_l=\ket{\psi'_l}\bra{\psi'_l}$ is an unnormalized pure state with
\begin{align}
    \ket{\psi'_l}=&\sum_{n=0}^\infty \frac{(t^2 \tanh{\mathcal{r}})^n}{\cosh{\mathcal{r}}}  \sum_{k_{min}}^l\sqrt{\binom{n}{k}\binom{n}{l-k}\binom{l}{k}} \nonumber \\
    &\times e^{ik\varphi}\ket{n+k-l}_a\ket{n-k}_b.
\end{align}
Eq.~\ref{eq:approx_povm_dm} reveals that even for values of $\eta$ deviating significantly from unity, we have a final state that is approximately pure, i.e., $\rho\approx \rho'_Z$ when $r^2(1-\eta) \ll 1$, since all terms with $l>Z$ in the sum can be neglected. This regime can always be reached by decreasing $r$ until the approximation holds. However, it is important to note that an imperfect detector reduces the success probability of the scheme by a factor of $\eta^{Z}$.
\subsection{Click/no-click detector}
In a similar manner to the above discussion, the PNR detector can be safely replaced with a click/no-click detector without ill-effect. Since we showed above that the effects of losses during state generation can be considered negligible if the beamsplitter reflectivity is tuned correctly, here we consider an ideal click detector with perfect efficiency. In the case where the detector reads no signal, then the POVM is simply the vacuum as before. In the case of a registered click, the POVM is given by
\begin{equation}
    D_{clk}=\sum_{l=1}^\infty \ket{l}\bra{l}.
\end{equation}

Now, the output density matrix is simply a mixture of all possible detected states with $Z\geq1$,
\begin{equation}
    \rho\propto\sum_{l=1}^\infty P(l)\rho_l,
    \label{eq:clk_noclk_dm}
\end{equation}
where each of the pure state in the mixture is weighted by the probability of the number of photons that actually went to the detector.
Since the quantities of interest are $\langle J'_z\rangle$ and $\Delta J'_z$, we can use the linearity of Eq.~\ref{eq:clk_noclk_dm} along with the results of Appx.~\ref{appx:MZI_calcs} to see that
\begin{align}
 \langle J'_z\rangle&=-\sin{\phi}\sum_{l=1}^\infty \text{Tr}\left[J_x \rho \right]\\
 &\propto \frac{-\sin{\phi}}{n}  
\end{align}
for each component state in the mixture, $\rho_l$. We also know that the maximum phase sensitivity is achieved about the angle of $\phi=0$, so the only necessary term from $\Delta J'_z$ is $\Delta J_z$. Thus, we can find that 
\begin{align}
    (\Delta J_z)^2 &= \sum_{l=1}^\infty \text{Tr}[J^2_z\rho]\\
    &\stackrel{(i)}{\leq} \sum_{l=1}^\infty\frac{l^2}{4}\left(\frac{r}{t}\right)^{2(l-1)},
\end{align}
which leads to a phase sensitivity about $\phi=0$ of
\begin{equation}
    \Delta \phi \propto \frac{ 1}{n}.
\end{equation}
These results can be easily understood by realizing that each of the possible pure state components in the mixture that results from the click detection has phase sensitivity scaling with the HL around \textit{the same} reference phase of $\phi=0$. Additionally, the high unbalancing of the beamsplitters makes the components in the mixture diminish with increasing $l$.

\begin{figure*}[ht!]
\centering
\includegraphics[width = 0.9\textwidth]{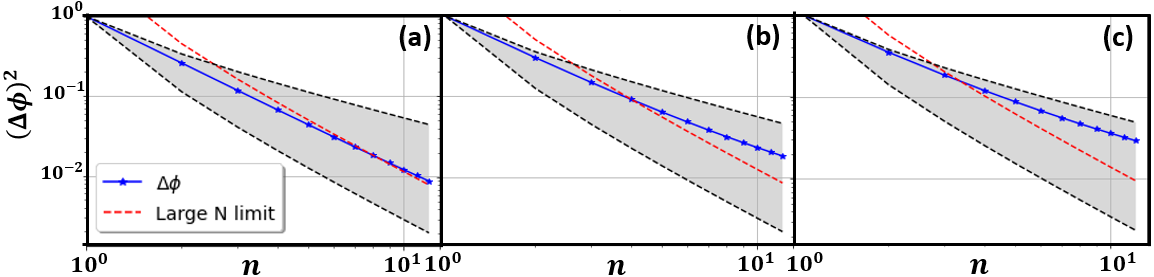}
\caption{Minimum phase sensitivity (blue) for a photon-subtracted state with beamsplitter reflectivities $r_1=r_2=0.1$ and input state $|n,n\rangle$, where the PNR detector has been replaced by an ideal click detector. Efficiency $\eta$ determines the combined transmission of the interferometer and end detector efficiencies and is $\eta=1$ (a), $\eta=0.95$ (b), $\eta=0.9$ (c). The shot-noise and HL bound the shaded region, and the red dotted line shows the Cram\'er-Rao bound for the large $N$ limit with $Z=1$ given by Eq.~\ref{eq:Dphi_i}.}
    \label{fig:phimin_n}
\end{figure*}

The results from this section are experimentally significant in that not only does the detector not need the ability to resolve photon-numbers, but the imperfect efficiency negligibly degrades the purity of the resultant state, provided that there is precise control in ensuring the beamsplitter reflectivity is small. 
\subsection{General OPO output}
The general output of an arbitrary twin-beam source, such as obtained from an above-threshold OPO, can be described as a mixture of the form 
\begin{align}
\rho &= \sum_{n,n'} \rho_{n,n'}\ket{n n}\bra{n' n'}.
\end{align}
Despite being a mixture, we show in Appx.~\ref{appx:gen_opo} that the indistinguishable multi-photon subtraction protocol also works for this input and leads to a Heisenberg-limited output with phase sensitivity scaling as $\Delta\phi \sim \langle N\rangle^{-1}$. The intuition behind this result follows similar reasons for why a click/no-click detector also fails to ruin Heisenberg-limited sensitivity; although the input is a mixture, each $|n_i,n_i\rangle$ component from the initial twin-beam is formed into a superposition with phase sensitivity of $\sim n_i^{-1}$ about the reference angle of $\phi=0$.  Since all of the components of the mixture are Heisenberg-limited at the same reference phase, then so is the entirety of the mixture. As such, when considering imperfections such as loss, simulations with the input state $|n,n\rangle$ can be reliably used to gauge to effectiveness of the process for an arbitrary twin-beam source with mean photon number $n$.

Fig.~\ref{fig:phimin_n} compares the phase sensitivity of a photon-subtracted state with an $|n,n\rangle$ input for the realistic scenario of a heralding PNR detector replaced by a click detector, and when the interferometer has losses. As shown in the previous section, inefficiencies in the click detector channel can be accounted for by choosing a small enough beamsplitter reflectivity, so the click detector here is considered ideal for simplicity. Additionally, if we assume that losses on both interferometer arms are identical, then the losses can be commuted and  combined with detection losses to result in a single overall efficiency of $\eta$ in each arm~\cite{oh2017practical}.

Each plot shows the phase sensitivity along with the large $n$ limit given by the Cram\'er-Rao bound for a single photon-subtraction given by Eq.~\ref{eq:Dphi_i} (red dotted line). When the beamsplitter reflectivities are small enough for the approximation to apply, as in Fig.~\ref{fig:phimin_n}a, it can be seen that the phase-sensitivity readily follows the Cram\'er-Rao bound, which scales with the HL up to a constant factor. This constant factor comes from the fact that since each mode in the twin beam has $n$ input photons, the total number of photons in the system is $2n$ leading to a HL of $(2n)^{-1}$; however, the ideal scaling with the single-photon subtracted state goes as $n^{-1}$. 
The second and third panels demonstrate that realistic interferometric losses do not drastically reduce the phase sensitivity, such as occurs with states less resilient to loss, such as $N00N$ states. As shown in Fig.~\ref{fig:phimin_n}b, the slope of the phase sensitivity still scales better than $\langle n\rangle^{-1/2}$ for $\eta=0.95$, and it is possible to achieve considerably larger overall $\eta$ in practical experiments with advanced low-loss optical coatings and highly efficient detectors~\cite{aLIGO2014}. Fig.~\ref{fig:phimin_n}c, with $\eta=0.9$, shows how larger losses on the order of $\langle N \rangle^{-1}$ bring the resultant phase-sensitivity away from the HL and back to scaling with the shot-noise, in agreement with the general result of Escher, de Matos Filho, and Davidovich~\cite{Escher2011}.

\section{Conclusion}

We have proposed and studied a nontrivial modification of the twin-beam input for Heisenberg-limited quantum interferometry, which features coherently indistinguishable multi-photon subtraction that leads to a superposition of photon-subtractions. This modification brings about a strong fringe signal --- absent from the unmodified twin-beam input --- while preserving Heisenberg-limited operation. The loss behavior is consistent with what is now well known about Heisenberg-limited interferometry. The experimental implementation should be feasible with state-of-the-art technology, for example using a stable OPO above threshold~\cite{Feng2004,Jing2006,Pysher2010a} and photodetectors with single-photon sensitivity. Detectors with PNR capability may improve the phase-sensitivity of the resultant photon-subtracted state by detecting $Z>1$, but this is not a requirement, and reasonable experimental losses still result in phase-sensitivity beating the standard quantum limit. We believe it is possible to operate at no more than $10^{6}$ photons per detection time bin, so as to be compatible with the lowest achievable optical losses and splitting ratios. 



\begin{acknowledgments}
We are grateful to Nicolas Treps, Claude Fabre, Luiz Davidovich, and Chun-Hung Chang for stimulating discussions. This work was supported by U.S. National Science Foundation grant PHY-1708023, the U.S. Defense Advanced Research Projects Agency, and an invited professorship to OP at Sorbonne Universit\'e. Additional support includes the Beitchman Award for Innovative Graduate Student Research in Physics in honor of Robert V. Coleman and Bascom S. Deaver, Jr.
\end{acknowledgments}

\newpage 
\begin{widetext}
\appendix

\section{Experimental State Derivation} \label{appx:exp_deriv}
Input state:
\begin{equation}
    \ket{\phi_{in}}=\ket{0,\mathcal{r}}_{ab}\ket{0}_c=\frac{1}{\cosh{\mathcal{r}}}\sum_n^\infty (\tanh{\mathcal{r}})^n \ket{n}_a\ket{n}_b\ket{0}_c
\end{equation}
The overall state is transformed by the beamsplitters and projective PNR measurement to become
\begin{equation}
    \ket{\phi_{out}}={}_c\bra{Z}U_{ac}U_{bc}\ket{\phi_{in}},
\end{equation}
where $U_{bc}=exp{\big[\theta_1(b c^\dag-b^\dag c)\big]}$, and $U_{ac}=exp\big[\theta_2(a c^\dag e^{-i\varphi}-a^\dag c e^{i\varphi})\big]$ to allow for an additional phase of $\varphi$ between mode $a$ and the reflected mode $c$ from the first beamsplitter, where reflectivities and transmissivities $r_i=\sin{\theta_i}$, $t_i=\cos{\theta_i}$. Note that the second beamsplitter operation, $U_{ac}$, will act between mode $a$ and the \textit{transformed} mode $c$ from the output of the first beamsplitter operation. With this in mind, the first beamsplitter transforms the input state to 
\begin{align}
    U_{bc}\ket{\phi_{in}}&=\sum_n^\infty\frac{(\tanh{\mathcal{r}})^n}{\sqrt{n!}\cosh{\mathcal{r}}}U_{bc}{b^\dag}^n U^\dag_{bc}\ket{n}_a\ket{0}_b\ket{0}_c \nonumber\\
    &=\sum_n^\infty\frac{(\tanh{\mathcal{r}})^n}{\sqrt{n!}\cosh{\mathcal{r}}}(t_1b^\dag+r_1c^\dag)^n\ket{n}_a\ket{0}_b\ket{0}_c \nonumber \\
    &=\sum_n^\infty\frac{(\tanh{\mathcal{r}})^n}{n!\cosh{\mathcal{r}}}\sum_k^n\binom{n}{k}t_1^k r_1^{n-k} {a^\dag}^n{b^\dag}^k{c^\dag}^{n-k}\ket{0}_a\ket{0}_b\ket{0}_c.
\end{align}
Now, applying the second beamsplitter yields the state
\begin{align}
    &\sum_n^\infty\frac{(\tanh{\mathcal{r}})^n}{n!\cosh{\mathcal{r}}}\sum_k^n\binom{n}{k}t_1^k r_1^{n-k} {b^\dag}^k(t_2a^\dag+e^{i\varphi}r_2c^\dag)^n(t_2c^\dag - e^{-i\varphi}r_2a^\dag)^{n-k}\ket{0}_{abc} \nonumber\\
    =&\sum_n^\infty\frac{(-r_1r_2^2\tanh{\mathcal{r}})^n}{n!\cosh{\mathcal{r}}}\sum_k^n\binom{n}{k}\left(\frac{t_1}{r_1r_2}\right)^k {b^\dag}^k\sum_j^n \binom{n}{j} \left(\frac{t_2}{r_2}\right)^{j} \nonumber\\
    & \times \sum_m^{n-k}\binom{n-k}{m}(-1)^{m+k}\left(\frac{t_2}{r_2}\right)^{m}(e^{i\varphi})^{k+m-j}{a^\dag}^{j+n-k-m}{c^\dag}^{n-j+m}\ket{0}_{abc}.
    \label{eq:appx_state_pre_pnr}
\end{align}
Projecting output mode $c$ onto the PNR detection event of $Z$ photons means that the only terms in the above sums that survive occur when $j=n+m-Z$. It is also important to note that the maximum value of $j$ is $n$, so the remaining sum over $m$ goes from zero to $m_{max}=\text{Min}(n-k,Z)$. These substitutions lead to
\begin{align}
    \ket{\phi_Z}=&\frac{r_2^Z\sqrt{Z!}}{t_2^Z\cosh{\mathcal{r}}}\sum^\infty_{n=0}(-t_2r_1r_2e^{-i\varphi}\tanh{\mathcal{r}})^n\sum^n_{k=0}\frac{\sqrt{(2n-k-Z)!}}{\sqrt{k!}}\left(\frac{-e^{i\varphi}t_1}{r_2r_2}\right)^k \nonumber\\
    &\times \sum^{m_{max}}_{m=0}\binom{n}{n+m-Z}\left(\frac{t_2}{r_2}\right)^{2m}\frac{(-1)^m}{m!(n-k-m)!}\ket{2n-k-Z}_a\ket{k}_b \nonumber\\
    \propto& \sum^\infty_{n=0}c_n\sum^n_{k=k_{min}}d_{n,k}\ket{n+k-Z}_a\ket{n-k}_b,
    \label{eq:appx_genresult}
\end{align}
where $k_{min}=\text{Max}(0,Z-n)$, $c_n=(t_1t_2\tanh{\mathcal{r}})^n$, and the coefficient $d_{n,k}$ is
\begin{equation}
        d_{n,k}=\frac{\sqrt{(n+k-Z)!}}{\sqrt{(n-k)!}}\left(\frac{r_1r_2}{-e^{i\varphi}t_1}\right)^k\sum_{m=0}^{\text{Min}(Z,k)}\binom{n}{n+m-Z}\left(\frac{t_2}{r_2}\right)^{2m}\frac{(-1)^m}{m!(k-m)!}. 
        \label{eq:appx_dn}
\end{equation}
The form of the general case in Eq.~\ref{eq:appx_genresult} shows that after all is said and done, we have a superposition over $n$ of two-mode superpositions that are desirable for quantum enhanced interferometry, where the $c_n$ terms depends on a new effective squeezing, which is reduced from the originial value by the transmissivity of the two beamsplitters. The success probability to create this state after a given $Z$ PNR detection is given by 
\begin{equation}
    P(Z)=\frac{Z!}{(\cosh{\mathcal{r}})^2}\left(\frac{r_2}{t_2}\right)^{2Z}\sum^\infty_{n=0}\sum^n_{k=k_{min}}c_n^2d_{n,k}^2
\end{equation}
\subsection{Highly unbalanced beamsplitters} \label{appx:unbal_BS}
 If we consider the case where both beamsplitters are identical and highly transmissive with $r_1=r_2\equiv r$ and $r\ll 1$, then examining Eq.~\ref{eq:appx_dn} shows that only the term with $m=k$ contributes to the coefficient $d_n$ to leading order.  Furthermore, since the sum over $m$ is truncated at $m_{max}=\text{Min}(k,Z)$, the sum over $k$ in Eq.~\ref{eq:appx_genresult} can be effectively truncated at $Z$ to the same order of approximation. The output state then becomes
 \begin{equation}
     \ket{\phi_Z}\propto\sum_{n=0}^\infty (t^2 \tanh{\mathcal{r}})^n  \sum_{k_{min}}^Z\sqrt{\binom{n}{k}\binom{n}{Z-k}\binom{Z}{k}}e^{ik\varphi}\ket{n+k-Z}_a\ket{n-k}_b,
     \label{eq:appx_approx_state}
\end{equation}
with an approximate success probability of 
\begin{equation}
    P(Z)\approx \frac{r^{2Z}}{(\cosh{\mathcal{r}})^2}\sum_{n=0}^\infty(1-2nr^2)(\tanh{\mathcal{r}})^{2n}\sum_{k_{min}}^Z\binom{n}{k}\binom{n}{Z-k}\binom{Z}{k} 
\end{equation}
\subsection{Coefficients}\label{appx:coef}
Here we verify several properties of the coefficients of the experimental state. In the case of a highly unbalanced beamsplitter, then for a given $n$, the experimental state takes the form of Eq.~\ref{eq:gen_coh_subtraction} with coefficients having the form shown in Eq.~\ref{eq:appx_approx_state} to be
\begin{equation}
    c_k\propto\sqrt{\binom{n}{k}\binom{n}{Z-k}\binom{Z}{k}}e^{ik\varphi},
\end{equation}
where all $c_k$ have the same proportionality constant from normalization. Setting $\varphi=0$ ensures that all $c_k$ are real. When writing the state in the Schwinger representation, these coefficients become
\begin{equation}
    c_m\propto\sqrt{\binom{n}{m+s}\binom{n}{s-m}\binom{2s}{m+s}},
    \label{eq:appx_m_coef}
\end{equation}
where $s=\tfrac{Z}{2}$ and $m=k-s$, from which it is clear to see that
\begin{equation}
    c_m=c_{-m}.  
\end{equation}
Now, how do neighboring coefficients relate? Calculating the ratio between neighbors yields
\begin{equation}
    \frac{c_{m+1}}{c_{m}}=\frac{s-m}{s+m+1}\left( \frac{n-s-m}{n-s+m+1}\right)^{1/2},
\end{equation}
which leads to the bounds of 
\begin{equation}
   \frac{1}{2s}\stackrel{(i)}{\leq}  \frac{c_{m+1}}{c_{m}}\stackrel{(i)}{\leq} 2s,
\end{equation}
where $(i)$ denotes the use of the approximation that $n\gg s$. This ratio is useful when calculating terms that appear in expectation values, and can be used to determine a bound on the sum of all pairs of neighboring coefficients to be
\begin{align}
     \frac{1}{2s}\stackrel{(i)}{\leq}\sum_{m=-s}^{s-1}c_{m}c_{m+1}\leq 1,
\end{align}
where the upper bound can be derived from the Cauchy–Schwarz inequality.
Similarly, the ratio between next neighboring coefficients has the bound
\begin{equation}
     \frac{c_{m+2}}{c_{m}}\stackrel{(i)}{\geq} \frac{1}{s(2s-1)},
\end{equation}
which leads to bounding the sum of next-nearest neighboring coefficients to be
\begin{equation}
   \frac{1}{2s^2-s} \stackrel{(i)}{\leq} \sum_{m=-s}^{s-2}c_{m}c_{m+2}\leq 1.
\end{equation}
\section{General Multi-Photon Subtracted State} \label{appx:gen_fisher}
Here we derive the Fisher information for the general case of the state given by Eq.~\ref{eq:gen_coh_subtraction}. All of the relevant terms are:
\begin{align}
    \langle  a^\dag b \rangle&= \sum_{k=0}^{Z-1}c_kc_{k+1}^*\sqrt{(n-k)(n-Z+k+1)} \nonumber \\
    \langle  a b^\dag \rangle&= \sum_{k=0}^{Z-1}c_k^*c_{k+1}\sqrt{(n-k+1)(n-Z+k)} \nonumber
        \\
        \langle  a^\dag a bb^\dag \rangle&= \sum_{k=0}^{Z}|c_k|^2(n-Z+k)(n-k+1)) \nonumber \\
    \langle  a a^\dag b^\dag b \rangle&= \sum_{k=0}^{Z}|c_k|^2(n-Z+k+1)(n-k)) \nonumber \\
    \langle  a^\dag a^\dag b b \rangle&= \sum_{k=0}^{Z-2}c_kc_{k+2}^*\sqrt{(n-k)(n-k-1)(n-Z+k+1)(n-Z+k+2)} \nonumber \\
    \langle a a b^\dag b^\dag \rangle&= \sum_{k=0}^{Z-2}c_k^*c_{k+2}\sqrt{(n-k+1)(n-k+2)(n-Z+k)(n-Z+k-1)}. \nonumber 
\end{align}
If we take a case where the number of photons removed from the state is small compared to the total, i.e., $n\gg Z$, then denoting the use of this approximation as $(i)$, the Fisher information is
\begin{align}
    \mathcal{F}&\stackrel{(i)}{=} 2\sum_{k=0}^Zn^2|c_k|^2- \sum_{k=0}^{Z-2}n^2(c_kc_{k+2}^*+c_k^*c_{k=2})+\left(\sum_{k=0}^{Z-1}nc_kc_{k+1}^*\right)^2 \nonumber \\
    &+\left(\sum_{k=0}^{Z-1}nc_k^*c_{k+1}\right)^2 -\sum_{k=0}^{Z-1}\sum_{k'=0}^{Z-1}n^2(c_kc_{k+1}^*c_{k'}^*c_{k'+1}+c_k^*c_{k+1}c_{k'}c_{k'+1}^*).
\end{align}
If we now make assumption $(ii)$ that all of the coefficients are the same, $c_k=\tfrac{1}{\sqrt{Z}}$, then
\begin{equation}
    \mathcal{F}\stackrel{(i,ii)}{=}\frac{4n^2}{Z},
\end{equation}
so the Cram\'er-Rao inequality leads to 
\begin{align}
    (\Delta \phi_d)^2& \geq \frac{Z}{N^2},
\end{align}
where $N=2n-Z$.
\subsection{MZI fringe and phase sensitivity}
\label{appx:MZI_calcs}
Using the Schwinger representation to determine the observables when the general Z-subtracted state from Eq.~\ref{eq:gen_coh_schw} is input into a MZI, we start by finding the expectation value of $J_x$ and $J_z$. We find that
\begin{align}
   \langle J_z\rangle &= 0  \\
   \langle J_x\rangle &= \frac12 \sum_{m=1-s}^{s}c_{m-1}^*c_{m}\sqrt{(j+m)(j-m+1)} +\frac12 \sum_{m=-s}^{s-1}c_{m+1}^*c_{m}\sqrt{(j-m)(j+m+1)} \nonumber \\
   &=\sum_{m=-s}^{s-1}c_{m+1}c_{m}\sqrt{(j-m)(j+m+1)} \nonumber \\
   &\stackrel{(i)}{=} j\sum_{m=-s}^{s-1}c_{m+1}c_{m}, \nonumber
\end{align}
where we have used that all $c_m \in \mathbb{R}$,  $c_m=c_{-m}$, and have used approximation $(i)$ that $j\gg s$. From here, we can estimate the value of the remaining sum by making use of the bounds on $\tfrac{c_{m+1}}{c_{m}}$, denoted by $(ii)$, for all of the coefficients derived in Appx~\ref{appx:coef}. This leads to the result that
\begin{align}
     \frac{j}{2s} \stackrel{(i,ii)}{\leq} \langle J_x\rangle &\stackrel{(i)}{\leq} j 
     \label{eq:appx_jx_ineq}
\end{align}
The observable fringe is given by the expectation value of $2J_z$ at the output, where
\begin{align}
    |\langle J_z'\rangle|&=|\cos{\phi} \langle J_z\rangle -\sin{\phi} \langle J_x\rangle| \nonumber\\
    \frac{j}{2s}|\sin{\phi}| &\stackrel{(i,ii)}{\leq} |\langle J_z'\rangle| \stackrel{(i)}{\leq} j|\sin{\phi}|.
    \label{eq:appx_coh_sub_fringe}
\end{align}
The end result above shows that the measurable fringe scales with the mean photon number of the state. When calculating $\Delta\phi$, the Heisenberg transformations yield
\begin{align}
\frac{d}{d\phi}\langle J_z' \rangle &= -\sin{\phi}\langle Jz\rangle - \cos{\phi}\langle Jx\rangle\\
(\Delta J_z')^2&=(\cos{\phi}\Delta J_z)^2+(\sin{\phi}\Delta J_x)^2 - \sin{\phi}\cos{\phi}(\langle\{J_z,J_x\}\rangle-2\langle Jz\rangle \langle Jx\rangle). 
\end{align}
Deriving the quantities individually, we have
\begin{align}
    (\Delta J_z)^2&= \langle J_z^2\rangle =\sum_{m=-s}^s m^2 |c_m|^2 
    < s^2,
    \label{eq:appx_jzvar}
\end{align}
where the inequality is obtained by replacing all $m^2$ with the maximimum value of $s^2$. We also find that
\begin{align}
 \langle J_x^2\rangle=&\,\frac14 \langle J_{+}^2+J_{-}^2 + 2(J^2 - J_{z}^2)\rangle \nonumber \\
 =&\, \frac12\sum_{m=-s}^s|c_m|^2(j^2+j-m^2)\nonumber \\
 &+\frac12\sum_{m=-s}^{s-2} c_{m}c_{m+2}\sqrt{(j+m+2)(j-m-1)(j+m+1)(j-m)} \nonumber\\
 \stackrel{(i)}{=}& \, \frac{j(j+1)}{2}-\frac{s^2}{2}+\frac12(j^2-s^2)\sum_{m=-s}^{s-2}c_mc_{m+2} \nonumber\\
 &\frac{j^2}{2}\left(1+\frac{1}{s(2s+1)}\right)+\frac{j}{2} \stackrel{(i,ii)}{\leq} \,  \langle J_x^2\rangle \stackrel{(i)}{\leq} 1,
\end{align}
where the sum in the second-to-last line was bounded by the ratio of next-nearest neighboring coefficients derived in Appx~\ref{appx:coef}. This leads to the variance
\begin{equation}
    (\Delta J_x)^2 \sim \mathcal{O}(j^2). 
    \label{eq:appx_jxvar}
\end{equation}
The other necessary terms are
\begin{align}
    \langle\{J_z,J_x\}\rangle&=\sum_{m=-s}^{s-1}(2m+1)c_{m+1}c_{m}\sqrt{(j-m)(j+m+1)}\approx \langle J_x \rangle \\
    2\langle Jz\rangle \langle Jx\rangle &=0,
\end{align}
which when combined with Eqs.~\ref{eq:appx_coh_sub_fringe}, \ref{eq:appx_jzvar}, and \ref{eq:appx_jxvar} lead to an overall value for the phase difference estimator given by Eq.~\ref{eq:dphi_MZI} to be
\begin{equation}
    (\Delta \phi)^2 \stackrel{(i)}{\leq} \frac{(s \cos{\phi})^2 + (\Delta J_x\sin{\phi})^2+\sin{\phi}\cos{\phi}\langle J_x \rangle}{(\tfrac{j}{2s}\cos{\phi})^2}
\end{equation}
This takes on a minimum value when $\phi=0$ to yield the upper bound
\begin{equation}
    \Delta \phi_{min} \stackrel{(i,ii)}{\leq} \frac{2s^2}{j},
\end{equation}
which scales with the HL up to a constant factor. By taking the upper limit of $\langle J_x \rangle$ from Eq.~\ref{eq:appx_jx_ineq}, the lower bound on $\Delta \phi$ when $\phi=0$ is
\begin{equation}
     \frac{\sqrt{s}}{j} \stackrel{(i,ii)}{\leq} \Delta \phi_{min}.
\end{equation}

\section{Phase sensitivity for a general twin-beam input}\label{appx:gen_opo}
In this section, we show that our photon subtraction protocol also works for the most general twin-beam statistical mixture, e.g.\ as produced by an OPO above threshold. 
The density operator in the Fock basis is given by
\begin{equation}
\label{eq:appx_gen_twin_beam}
\rho_{ab} = \sum_{n,n'} \rho_{n,n'}\ket{n n}\bra{n' n'} \ .
\end{equation}
The two beamplitter operations are given by
\begin{align}
U_{ac'}U_{bc}&=exp\left[\theta_2(a^\dag c'-ac'^\dag )\right] exp\left[\theta_1(b^\dag c-bc^\dag)\right] \nonumber \\
&=\sum_{j,k}\frac{\theta_1^k\theta_2^j}{j!k!}\left(a^\dag c'-ac'^\dag \right)^j\left(b^\dag-bc^\dag\right)^k,
\end{align}
where $c'=c\cos{\theta_1}-b\sin{\theta_1}$ is the transformed vacuum mode from the first beamsplitter input. Because the input state, $\rho_{ab}$, consists solely of vacuum in the input mode $c$, and we are post-selecting the transformed mode $c$ on a detection of $Z$ photons, we need only consider terms with of the form $c^x c^{\dag(x+Z)}$ and ${c}^{\dag (x+Z)}c^x$. To further simplify, we can assume the highly unbalanaced beamsplitter regime, where both $\theta_1 \ll 1$ and $\theta_2 \ll 1$, in which case we need only consider terms with $c^{\dag Z}$. Thus, to leading order in $\theta$, we have $j+k=Z$ and
\begin{equation}
    U_{ac'}U_{bc}\approx \sum_{k=0}^Z \frac{r_1^k r_2^{Z-k}}{k!(Z-k!)}\left(-at_1\right)^{Z-k}\left(-b\right)^k c^{\dag Z},
\end{equation}
 where $r_1=\sin{\theta_1}\approx\theta_1$ and $r_2=\sin{\theta_2}\approx\theta_2$. Sending the twin-beam input through the unbalanced beamsplitters and detecting $Z$ photons in mode $c$ leads to
 \begin{align}
     \rho_{out}=&\text{Tr}_c\left[(|Z\rangle\langle Z|)_{c} U_{ac'}U_{bc}\rho_{ab}\otimes|0\rangle \langle0|U^\dag_{bc}U^\dag_{ac'}\right]\\
     =&\text{Tr}_c\left[(|Z\rangle\langle Z|)_{c} \sum^Z_{k,k'}\frac{r_1^{k+k'}r_2^{2Z-k-k'}t_1^{2Z-k-k'}}{k!k'!(Z-k)!(Z-k')!}\sum^\infty_{n,n'}\rho_{n,n'}a^{Z-k}b^kc^{\dag Z}|n, n, 0\rangle \langle n',n',0|a^{\dag(Z-k')}b^{\dag k'}c^Z\right]\\
     \propto&(r_2t_1)^{2Z}\sum^\infty_{n,n'}\rho_{n,n'}\sum^Z_{k,k'}C_{k,k'}|n-Z+k,n-k\rangle\langle n'-Z+k',n'-k'|,
 \end{align}
 where $C_{n,n'}$ contains all of the remaining binomial coefficients,
 \begin{equation}
     C_{n,n'}=\left[\binom{n}{k}\binom{Z}{k}\binom{n}{Z-k}\binom{n'}{k'}\binom{Z}{k'}\binom{n'}{Z-k'}\right]^{1/2}\left(\frac{r_1}{t_1r_2}\right)^{k+k'}.
 \end{equation}
 Writing the output in the Schwinger representation, we have
 \begin{equation}
     \rho_{out}=\sum^\infty_{j,j'}\rho_{j,j'}\sum^s_{m,m'=-s}C'_{m,m'}|j,m\rangle\langle j',m'|,
 \end{equation}
 where $s=\tfrac{Z}{2}$. From here, the calculations for the Schwinger operators follow the form of \ref{appx:MZI_calcs} for each of the superpositions within the mixture, leading to the finding that about the interferometric phase $\phi=0$, we have that $(\Delta J_z')^2 \leq s^2$. Additionally, following the arguments of \ref{appx:coef}, we find that $\langle J_x \rangle\geq \tfrac{j_{avg}}{2s}$, where $j_{avg}$ is the average value of $j$ in the statistical mixture. These values lead to the determination that about $\phi=0$,
 \begin{equation}
     \Delta\phi=\frac{(\Delta J_z')}{\left|\frac{d}{d\phi}\langle J_z' \rangle\right|}|_{\phi=0}\stackrel{(i,ii)}{\leq} \frac{2s^2}{j_{avg}},
 \end{equation}
and hence the general twin-beam source is sufficient to achieve phase-sensitivity scaling proportionally with the HL.
\end{widetext}


%

\end{document}